\documentclass{article}

\usepackage{authblk}
\usepackage[utf8]{inputenc} 
\usepackage[T1]{fontenc}    
\usepackage{hyperref}       
\usepackage{url}            
\usepackage{booktabs}       
\usepackage{amsfonts}       
\usepackage{amsmath,amssymb}
\usepackage{nicefrac}       
\usepackage{microtype}      
\usepackage{graphicx}
\usepackage{color}
\usepackage[sorting=none]{biblatex}
\usepackage{makecell}

\def\qualifigsep{.1mm}

\title{XPDNet for MRI Reconstruction: an application to the 2020 fastMRI challenge}

\author[1]{Zaccharie Ramzi}
\author[2]{Jean-Luc Starck}
\author[3]{Philippe Ciuciu}
\affil[1]{CEA (Neurospin \& Cosmostat), Inria (Parietal)}
\affil[2]{AIM, CEA, CNRS, Université Paris-Saclay, Université Paris Diderot, Sorbonne Paris Cité}
\affil[3]{Neurospin, Inria (Parietal)}

\date{November 2020}

\begin{document}

\maketitle

\section{Synopsis}
We present a new neural network, the XPDNet, for MRI reconstruction from periodically under-sampled multi-coil data. 
We inform the design of this network by taking best practices from MRI reconstruction and computer vision.
We show that this network can achieve state-of-the-art reconstruction results, as shown by its ranking of second in the fastMRI 2020 challenge.

\section{Main findings}
We introduce the XPDNet, a neural network for MRI reconstruction, which ranked second in the fastMRI 2020 challenge.


\section{Introduction}

The acceleration of the scan time of Magnetic Resonance Imaging (MRI) can be done using Compressed Sensing (CS)~\cite{Lustig2007} and Parallel Imaging (PI)~\cite{Griswold2002GeneralizedGRAPPA, Pruessmann1999SENSE:MRI}.
To have access to the underlying anatomical object $x\in\mathbb{C}^{n \times n}$, one must solve the following idealized inverse problem:
\begin{equation}
    \label{eq:ideal_inv-pb}
    M_{\Omega} F S_\ell x = y_\ell, \quad \forall \ell=1,\ldots, L
\end{equation}
where $M_{\Omega}\in\{0,1\}^{m\times n} $ is a mask indicating which points in the Fourier space (also called k-space) are sampled, $S_\ell\in\mathbb{C}^{n\times n}$ is a sensitivity map for coil $\ell$, $F$ is the 2D Fourier transform, and $y_\ell\in\mathbb{C}^m$ is the k-space measurement for coil $\ell$. Problem~\eqref{eq:ideal_inv-pb} is typically solved with optimization algorithms that introduce a regularisation term to cope with its indeterminacy.

\begin{equation}
    \underset{x \in \mathbb{C}^{n}}{\textcolor{blue}{\operatorname{arg\,min}}} \sum_{\ell=1}^L \frac{1}{2} \textcolor{blue}{\|}y_\ell-M_{\Omega} \textcolor{blue}{S_\ell} x\textcolor{blue}{\|_{2}^{2}}+\textcolor{blue}{R(x)}
    \label{eq:opt-prob}
\end{equation}

In Eq.~\eqref{eq:opt-prob}, the terms in color reflect parts of this optimization that are, to some degree, handcrafted and could benefit from learning.
In plain text, the places where learning could improve the quality and the speed of the reconstruction are:
\begin{itemize}
    \item the optimization algorithm hyper-parameters (symbolized by $\operatorname{arg\,min}$) like step sizes and momentum;
    \item the sensitivity maps $(S_{\ell})_\ell$, usually estimated using ESPIRiT~\cite{Uecker2014ESPIRiT-AnAccess};
    \item the regularizer $R(x)$, usually $\lambda \|\psi x\|_1$ where $\psi$ is the decomposition over some wavelet basis;
    \item the data consistency term $\sum_{\ell=1}^L \frac{1}{2}\left\|y_{\ell}-M_{\Omega} S_{\ell} x\right\|_{2}^{2}$, usually derived from an additive white Gaussian noise model.
\end{itemize}

\emph{XPDNet}, the neural network introduced in this abstract, leverages all the room for learning to provide an efficient, i.e. fast and accurate, MRI reconstruction method.
 
\section{Methods}

\begin{figure}[h]
\centering
\includegraphics[width=0.8\textwidth]{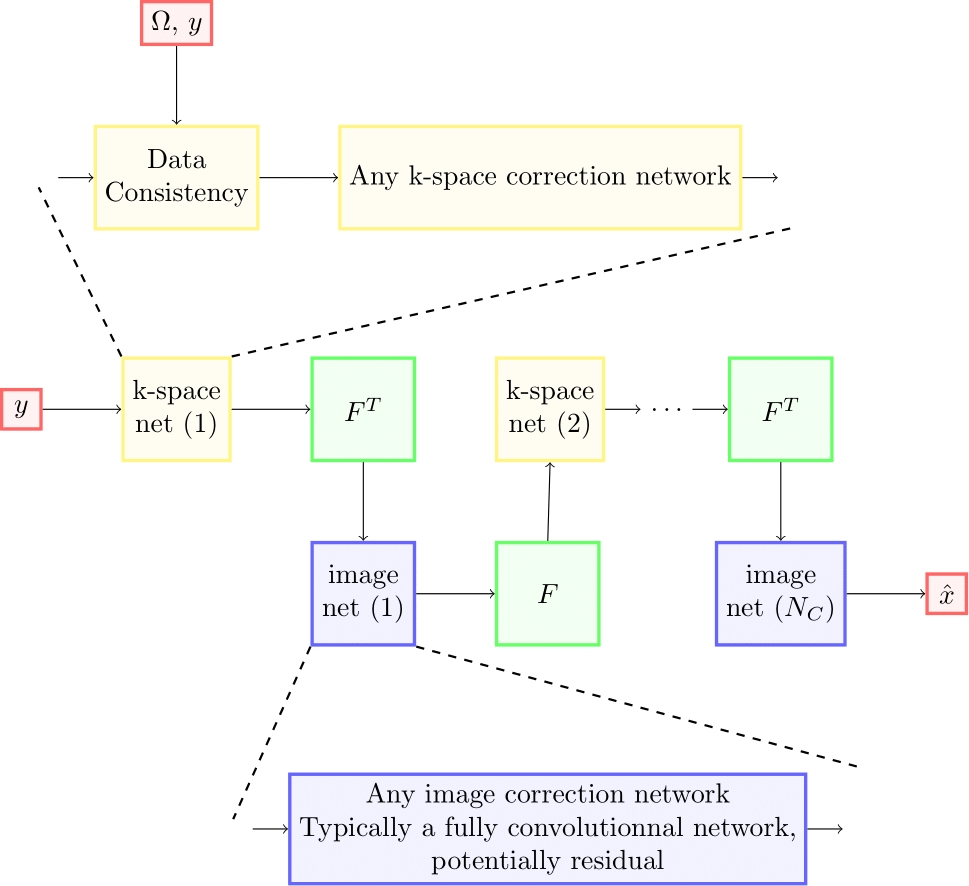}
\caption{
\label{fig:cd-net} 
General cross-domain networks architecture. Skip and residual connection are omitted for the sake of clarity. $y$ are the under-sampled measurements, in our case the k-space measurements, $\Omega$ is the under-sampling scheme, $F$ is the measurement operator, in our case the Fourier Transform, and $\hat{x}$ is the recovered solution.
}
\end{figure}

\paragraph{Cross-domain networks.}
The general intuition behind cross-domain networks is that we are going to alternate the correction between the image space and the measurements space.
The key tool for that is the unrolling of optimisation algorithms introduced in~\cite{Gregor2010}.
An illustration of what cross-domain networks generally look like is provided in Fig.~\ref{fig:cd-net}.

\paragraph{Unrolling the PDHG.}The \emph{XPDNet} is a particular instance of cross-domain networks.
It is inspired by the PDNet introduced in~\cite{Adler2018} by unrolling the Primal Dual Hybrid Gradient~(PDHG) algorithm~\cite{Chambolle2011}.
In particular, a main feature of the PDNet is its ability to learn the optimisation parameters using a buffer of iterates, here of size 5.

\paragraph{Image correction network.} The plain Convolutional Neural Network (CNN) is replaced by an Multi-scale Wavelet CNN (MWCNN)~\cite{Liu}, but the code\footnote{\href{https://github.com/zaccharieramzi/fastmri-reproducible-benchmark}{github.com/zaccharieramzi/fastmri-reproducible-benchmark}} allows for it to be any denoiser, hence the presence of X in its name.
We chose to use a smaller image correction network than that presented in the original paper~\cite{Liu}, in order to afford more unrolled iterations in memory~\cite{Ramzi2020BenchmarkingDatasets}.

\paragraph{k-space.} In this challenge, since the data is multicoil, we did not use any k-space correction network which would be very demanding in terms of memory footprint.
However, following the idea of~\cite{Sriram2020End-to-EndReconstruction}, we introduced a refinement network for $(S_\ell)_\ell$, initially estimated from the lower frequencies of the retrospectively under-sampled coil measurements.
This sensitivity maps refiner~\cite{Sriram2020End-to-EndReconstruction} is chosen to be a simple U-net~\cite{Ronneberger}.

We therefore have 25 unrolled iterations, an MWCNN that has twice as less filters in each scale, a sensitivity maps refiner smaller than that of~\cite{Sriram2020End-to-EndReconstruction} and no k-space correction network.

\paragraph{Training details.} The loss used for the network training was a compound loss introduced by~\cite{Pezzotti2020AnChallenge}, consisting of a weighted sum of an $L_1$ loss and the multi-scale SSIM (MSSIM)~\cite{Wang2004}.
The optimizer was the Rectified ADAM (RAdam)~\cite{Liu2020OnBeyond} with default parameters\footnote{\href{https://www.tensorflow.org/addons/api_docs/python/tfa/optimizers/RectifiedAdam}{tensorflow.org/addons/api\_docs/python/tfa/optimizers/RectifiedAdam}}.
The training was carried for 100 epochs~(batch size of 1) and separately for acceleration factors 4 and 8. 
The networks were then fine-tuned for each contrast for 10 epochs.
On a single V100 GPU, the training lasted 1 week for each acceleration.

\paragraph{Data.} The network was trained on the brain part of the fastMRI dataset~\cite{Zbontar}.
The training set consists of 4,469 volumes from 4 different contrasts: T1, T2, FLAIR and T1 with admissions of contrast agent~(labelled T1POST). The validation was carried over 30 contrast-specific volumes from the validation set.

\section{Results}

\begin{figure}[h]
\begin{center}
\hspace*{-4.7cm}
\begin{tabular}{c@{\hspace*{\qualifigsep}}c@{\hspace*{\qualifigsep}}c@{\hspace*{\qualifigsep}}c@{\hspace*{\qualifigsep}}c}
\makecell{{\bf T1} \\ PSNR: 41.56 \\ SSIM: 0.9506} & \makecell{{\bf T2} \\ PSNR: 40.68 \\ SSIM: 0.9554} & \makecell{{\bf FLAIR} \\ PSNR: 39.60 \\ SSIM: 0.9321} & \makecell{{\bf T1POST} \\ PSNR: 42.53 \\ SSIM: 0.9683} \\
\includegraphics[scale=0.47]{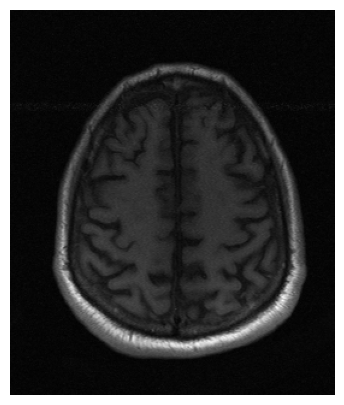}&
\includegraphics[scale=0.47]{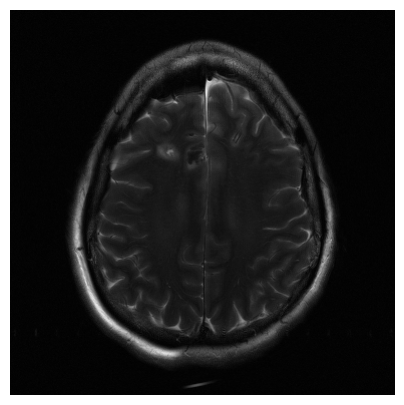}&
\includegraphics[scale=0.47]{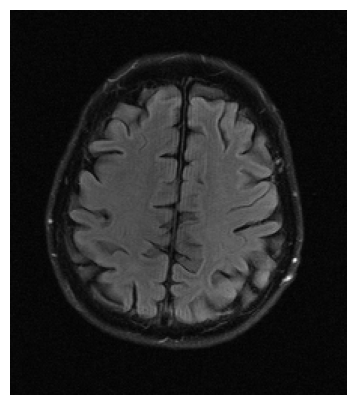}&
\includegraphics[scale=0.47]{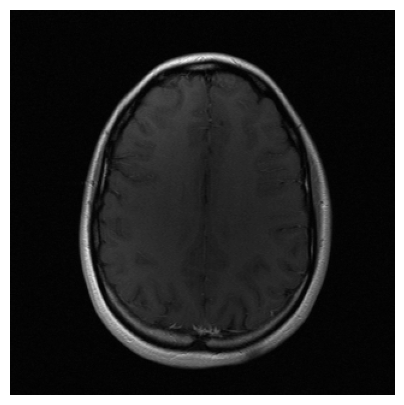}\\
\includegraphics[scale=0.47]{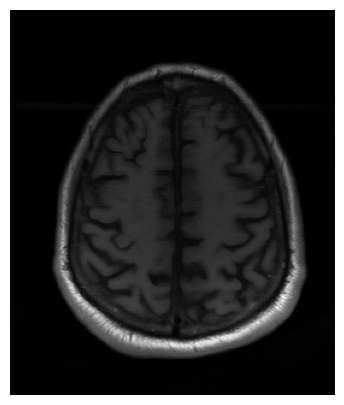}&
\includegraphics[scale=0.47]{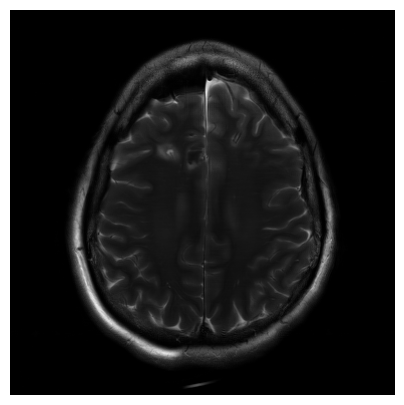}&
\includegraphics[scale=0.47]{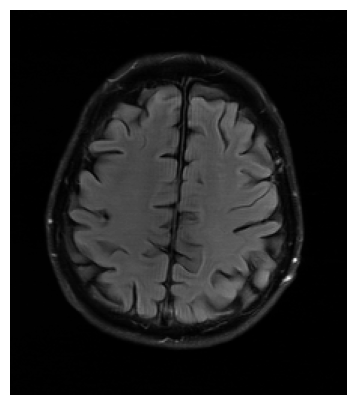}&
\includegraphics[scale=0.47]{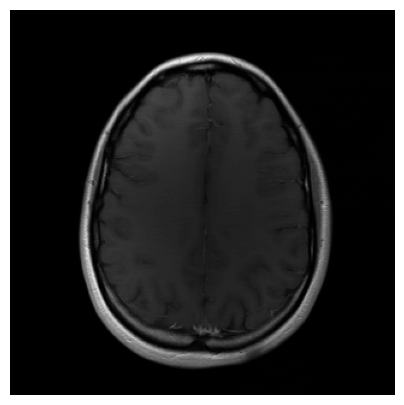}\\
\includegraphics[scale=0.47]{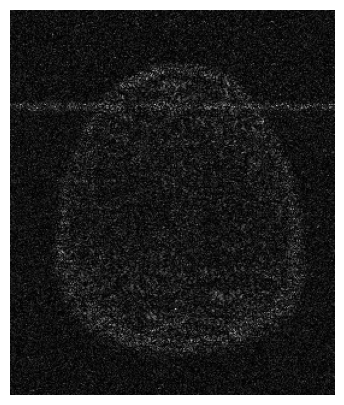}&
\includegraphics[scale=0.47]{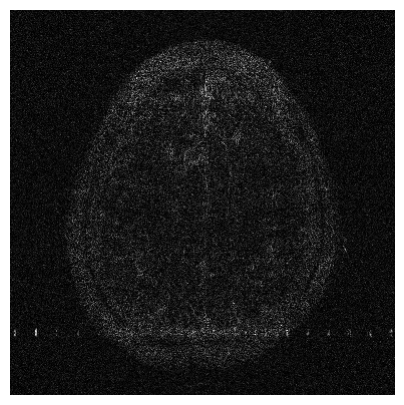}&
\includegraphics[scale=0.47]{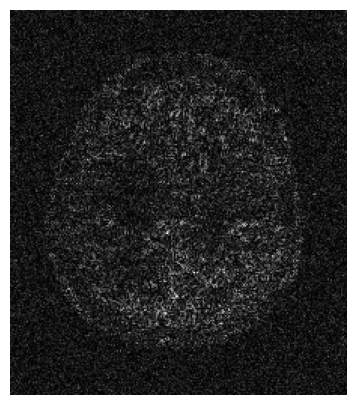}&
\includegraphics[scale=0.47]{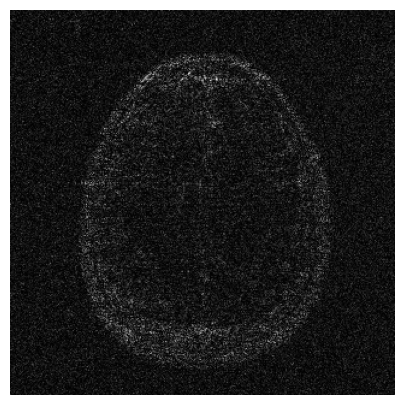}
\end{tabular}
\caption{Magnitude reconstruction results for the different fastMRI contrasts at acceleration factor 4. The top row represents the ground truth, the middle on represents the reconstruction from a retrospectively under-sampled k-space, and the bottom row represents the absolute error when comparing the two. The average image quantitative metrics are given for 30 validation volumes. \label{fig:res-af4}}
\end{center}
\end{figure}

\begin{figure*}[h]
\begin{center}
\hspace*{-4.7cm}\begin{tabular}{c@{\hspace*{\qualifigsep}}c@{\hspace*{\qualifigsep}}c@{\hspace*{\qualifigsep}}c@{\hspace*{\qualifigsep}}c}
\makecell{{\bf T1} \\ PSNR: 38.57 \\ SSIM: 0.9348} & \makecell{{\bf T2} \\ PSNR: 37.41 \\ SSIM: 0.9404} & \makecell{{\bf FLAIR} \\ PSNR: 36.81 \\ SSIM: 0.9086} & \makecell{{\bf T1POST} \\ PSNR: 38.90 \\ SSIM: 0.9517} \\
\includegraphics[scale=0.47]{Figures/imageAXT1_gt.png}&
\includegraphics[scale=0.47]{Figures/imageAXT2_gt.png}&
\includegraphics[scale=0.47]{Figures/imageAXFLAIR_gt.png}&
\includegraphics[scale=0.47]{Figures/imageAXT1POST_gt.png}\\
\includegraphics[scale=0.47]{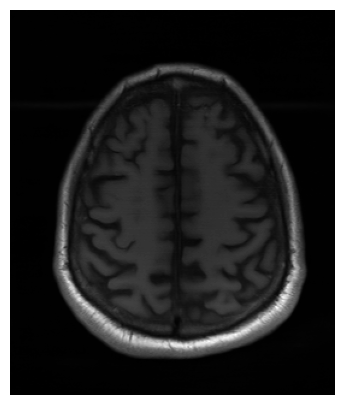}&
\includegraphics[scale=0.47]{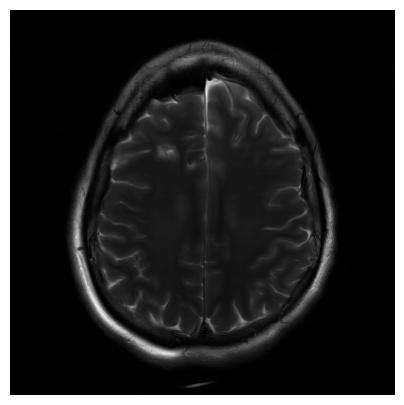}&
\includegraphics[scale=0.47]{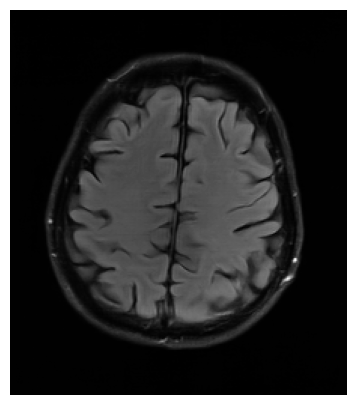}&
\includegraphics[scale=0.47]{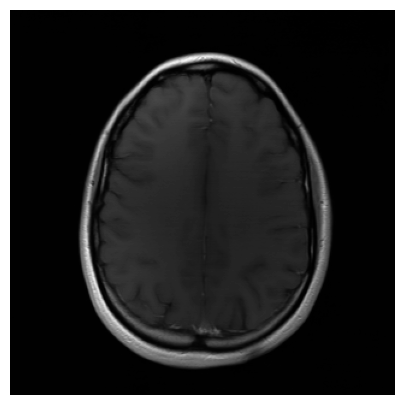}\\
\includegraphics[scale=0.47]{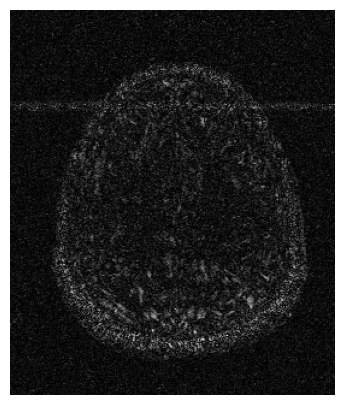}&
\includegraphics[scale=0.47]{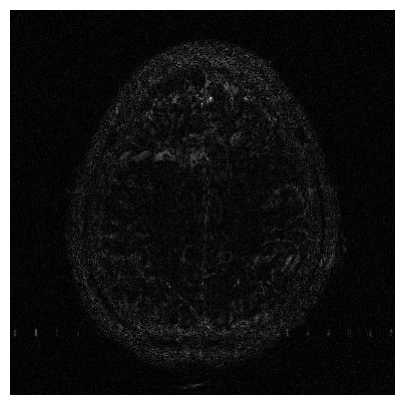}&
\includegraphics[scale=0.47]{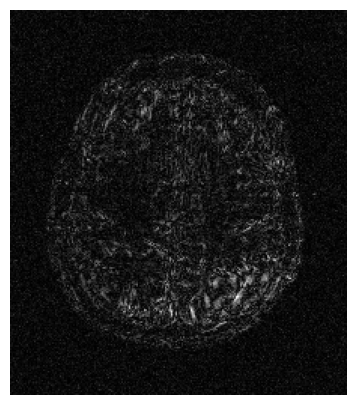}&
\includegraphics[scale=0.47]{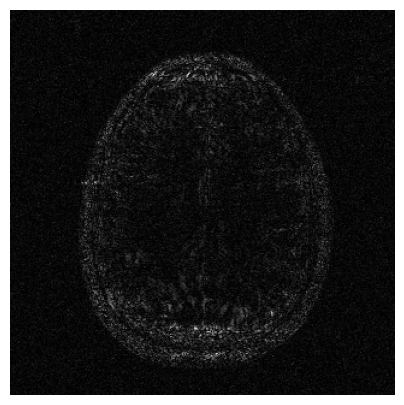}
\end{tabular}
\caption{Magnitude reconstruction results for the different fastMRI contrasts at acceleration factor 8. The top row represents the ground truth, the middle one represents the reconstruction from a retrospectively under-sampled k-space, and the bottom row represents the absolute error when comparing the two. The average image quantitative metrics are given for 30 validation volumes. \label{fig:res-af8}}
\end{center}
\end{figure*}

\paragraph{Quantitative.}
We used the PSNR and SSIM metrics to quantitatively compare the reconstructed magnitude image and the ground truth.
They are given for each contrast and for the 2 acceleration factors in the Figs.~\ref{fig:res-af4}-~\ref{fig:res-af8}.
Similar results are available on the public fastMRI leaderboard\footnote{\href{https://fastmri.org/leaderboards/}{fastmri.org/leaderboards}}, although generally slightly better.
It is a bit difficult to consider these results when compared to only the zero-filled metrics, but these quantitative metrics do not accurately capture the performance of the GRAPPA algorithm~\cite{Griswold2002GeneralizedGRAPPA}.
However, at the time of submission, this approach ranks 2nd in the fastMRI leaderboards for the PSNR metric, and finished 2nd in the 4$\times$ and 8$\times$ tracks of the fastMRI 2020 brain reconstruction challenge~\cite{Muckley2020State-of-the-artChallenge}.

\paragraph{Qualitative.}
The visual inspection of the images reconstructed~(available in Fig.~\ref{fig:res-af4}) at acceleration factor 4 shows little to no visible difference with the ground truth original image.
However, when increasing the acceleration factor to 8, we can see that smoothing starts to appear which leads to a loss of structure as can be seen in Fig.~\ref{fig:res-af8}.

\section{Conclusion and Discussion}
We managed to gather insights from many different works on computer vision and MRI reconstruction to build a modular approach.
Currently our solution XPDNet is among the best in PSNR and NMSE for both the multicoil knee and brain tracks at the acceleration factors 4 and 8.
Furthermore, the modularity of the current architecture allows to use the newest denoising architectures when they become available.
However, the fact that this approach fails to outperform the others on the SSIM metric is to be investigated in further work.

\section{Figures}

\medskip


\printbibliography

\end{document}